\documentclass[11pt,twoside]{article}
\usepackage{asp2004}
\usepackage{natbib}
\usepackage{psfig}
\usepackage{epsf}
\usepackage{graphics}
\usepackage{lscape}
\def\w{{\bf w\ }}

\begin{document}
\title{Cluster Survey Studies of the Dark Energy}
\author{Joseph J. Mohr}
\affil{Department of Astronomy and Department of Physics\\
1002 W. Green St, University of Illinois, Urbana, IL}

\begin{abstract}
Galaxy cluster surveys are power tools for studying the dark energy.  In principle, the equation of state parameter \w of the dark energy and its time evolution can be extracted from large solid angle, high yield surveys that deliver tens of thousands of clusters.  Robust constraints require accurate knowledge of the survey selection, and crude cluster redshift estimates must be available.  A simple survey observable like the cluster flux is connected to the underlying cluster halo mass through a so--called mass--observable relation.  The calibration of this mass--observable relation and its redshift evolution is a key challenge in extracting precise cosmological constraints.  Cluster survey {\it self--calibration} is a technique for meeting this challenge, and it can be applied to large solid angle surveys.  In essence, the cluster redshift distribution, the cluster power spectrum, and a limited number of mass measurements can be brought together to calibrate the survey and study the dark energy simultaneously.  Additional survey information like the shape of the mass function and its evolution with redshift can then be used to test the robustness of the dark energy constraints.
\end{abstract}
\thispagestyle{plain}

\section*{Introduction to Cluster Surveys}
The observed cosmic acceleration raises fundamental questions about the expanding universe and our understanding of gravity.  It should come as no surprise that scientists are investing time in developing tools to study this cosmic acceleration.  The leading tools currently
include distance measurements using SNe Ia, weak gravitational lensing studies of the matter distribution, galaxy power spectrum studies, and galaxy cluster surveys to quantify the emergence and evolution of massive objects in the universe.  Published constraints on the nature of the dark energy that rely on combinations of existing datasets are already quite interesting \citep{spergel03,tegmark04,seljak04}.  A task for scientists is to construct far more sensitive new experiments that employ each of these techniques individually to provide tight and independent constraints on the dark energy equation of state parameter $w$.  Only then can we make meaningful tests of consistency among the experiments; the independent results can be combined with some confidence to provide mixed--technique constraints on \w and its redshift evolution with dramatically improved precision. 

The cluster survey technique provides leverage on the dark energy through the dark energy sensitivity of the expansion history of the universe and the growth rate of structures in our expanding universe \citep{haiman01}.  The cluster redshift distribution ($d^2N/dzd\Omega$: number of clusters per unit redshift per unit solid angle) is the product of the volume surveyed $d^2V/dzd\Omega$ and the number density or abundance of clusters $n(z)$.  The volume surveyed per unit redshift and solid angle depends on the angular diameter distance $d_A(z)$ and the Hubble parameter $H(z)$ at that redshift.  The abundance is an integral over the mass function  of clusters ($dn/dM$: number density per unit mass) times a selection function $f(M,z)$, which describes the sensitivity of the survey to clusters of mass $M$ at redshift $z$.
\begin{equation}
{d^2N\over dz\,d\Omega}={c\over H(z)} d^2_A(z)\left(1+z\right)^2 \,\, \int_0^\infty dM\, {dn\over dM}(M,z)\, f(M,z)
\end{equation}
The evolution of the halo mass function $dn/dM$ can be understood within the context of the linear growth of density perturbations \citep[i.e.][]{press74,sheth99}. This evolution is now well studied using N--body simulations, and further improvements are expected \citep[e.g.][]{jenkins01,hu03a}.  

\begin{figure}[!ht] 
\plotfiddle{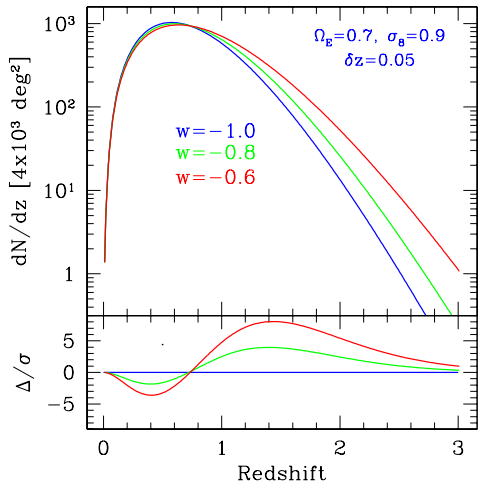}{2.3in}{0}{35}{35}{-110}{-60}
\caption{The expected redshift distribution (top) and quantified differences (bottom) for models where only the dark energy equation of state parameter \w is varied.  These models are all normalized to produce the same local abundance of galaxy clusters.  Note that for models with higher \w there is less volume, reducing the number of clusters at $z\sim0.5$, and structures grow less rapidly, increasing the number of clusters at higher redshifts.
\label{fig:wdepend}}
\end{figure} 

The sensitivity of the cluster redshift distribution to \w is illustrated in Figure~\ref{fig:wdepend}.  Three redshift distributions for a large Sunyaev--Zel'dovich Effect (SZE) cluster survey are shown; each corresponds to a different value of $w$.  The statistical significance of the differences among the models appears in the lower panel.  The volume sensitivity dominates at intermediate redshifts $z\sim0.5$, and the growth rate sensitivity dominates at high redshift.  

For large solid angle cluster surveys there are additional observables, including the clustering power spectrum for the galaxy clusters \citep{majumdar04,lima04}.  Galaxy clusters, like other cosmic objects, act as tracers of the underlying dark matter.  In the case of galaxy clusters, the most massive collapsed objects in the universe, the exact relationship between the cluster power spectrum and the dark matter power spectrum is well understood theoretically \citep[i.e.][]{mo96}, and this relationship or biasing is a function of cluster mass.  Thus, measurements of the cluster power spectrum provide important additional information about cosmology and cluster masses.  Moreover, the distribution of clusters in observed flux at each redshift provides additional cosmological and cluster mass information \citep{hu03b}.  Finally, direct mass measurements-- through X-ray observations, optical spectroscopy or weak lensing-- provide important additional leverage on cosmology and cluster masses \citep{majumdar03,majumdar04}.  Combining this myriad of observables is discussed in the last section of this proceeding.

\section*{Anatomy of Several Large Cluster Surveys}

There are several requirements, which must be in place to achieve precise cosmological constraints from a cluster survey:  (a) a firm theoretical understanding of the formation and evolution of massive dark matter halos in our expanding universe (as described above), (b) a clean and well--understood selection of  large numbers ($\sim10^4$) of massive dark matter halos (i.e. galaxy clusters) over a range of redshifts, (c)   a cluster survey observable that correlates with the cluster halo mass (this is the so-called mass--observable relation, and many have been published in the literature; see last section of this proceeding), and (d) crude redshift estimates for each cluster (only plausible with photometric redshifts).

There are many cluster surveys underway or planned in the X--ray, SZE, optical and NIR.  Only a few of these surveys are of large, contiguous solid angle with the sensitivity to probe to redshift $z\sim1$.  Ongoing surveys include the Red Sequence Cluster Survey (RCS II), which is a two band optical survey being carried out on the Canada--France--Hawaii Telescope, and the Massive Abell Cluster survey (MACS), which is a low yield X--ray survey utilizing the ROSAT all sky survey.  Funded surveys in the build phase include the South Pole Telescope (SPT; http://spt.uchicago.edu) Sunyaev-Zel'dovich Effect (SZE) survey (first light 2007) and the {\it Planck} Surveyor SZE survey (first light 2008).  The Dark Universe Observatory (DUO; http://duo.gsfc.nasa.gov) is a dedicated X-ray survey mission at the end of Phase A study (launch in 2007), and the Dark Energy Survey (DES; http://cosmology.astro.uiuc.edu/DES) is a deep, multiband optical survey (currently partially funded; first light in 2009) coordinated with SPT.  Even more aggressive missions that are expected to come online in about a decade include the Large Synoptic Survey Telescope (LSST; first light 2012), which is a time domain survey of about half the sky, and the Supernova Acceleration Probe (SNAP; launch in 2014), which is slated to include a large survey component.  In addition to these massive cluster surveys, there are many smaller efforts that will serve important roles as science and technology precursors, but in my view it is unlikely that these projects will deliver the cluster numbers or survey solid angle required to to self--calibrate and achieve robust constraints on the dark energy (barring a major breakthrough in robust and direct cluster mass measurements out to redshifts $z\sim1$).  These include serendipitous X--ray surveys using the {\it Chandra} and XMM--{\it Newton} archives, near term SZE surveys like the Sunyaev-Zel'dovich effect Array (SZA), the Arcminute Microkelvin Imager (AMI), the Atacama Pathfinder Experiment (APEX), and deep near-infrared and optical surveys like the FLAMINGOS extragalactic survey (FLAMEX).

It's interesting to note that people are pursuing cluster cosmology using a wide range of cluster observables, including the X--ray emission from the intracluster medium (ICM), the effects the ICM has on the spectrum of the cosmic microwave background (SZE), the light and number of cluster galaxies, and the weak lensing shear signature of the clusters.  Generally speaking, cluster selection is cleanest and cluster mass--observable relations are tightest in the X-ray; there is a strong theoretical prejudice (supported in hydrodynamical simulations) that the situation will be even better at mm wavelengths (i.e. SZE observations).  Cluster finding that employs cluster galaxies is improving rapidly \citep[i.e.][]{gladders00,gladders02,bahcall03}, but the optical or near-infrared mass--observable relations in the literature \citep[i.e.][]{kochanek03,lin03b,yee03,lin04a} tend to have a factor of 3 to 4 larger  scatter (e.g. the fractional accuracy in the mass estimate from a given observable) than similar, published analyses in the X--ray \citep[i.e.][]{mohr99,finoguenov01,reiprich02}.    Weak lensing surveys are potentially very powerful, but only for finding very massive clusters, where the large scale structures projected along the line of sight do not swamp the cluster shear signal \citep{metzler01,dodelson04}.   Stacked shear profiles of large numbers of low mass clusters should be important for mass--calibration of all surveys.  It is important to realize that photometric redshifts for the clusters require a multiband, optical survey, and so every large cluster survey will have an extensive optical dataset.  The field is developing rapidly, and there's every reason to believe new techniques will be introduced that make improved use of the less well studied galaxy, weak lensing and SZE properties of clusters.

\begin{figure}[!ht] 
\plotfiddle{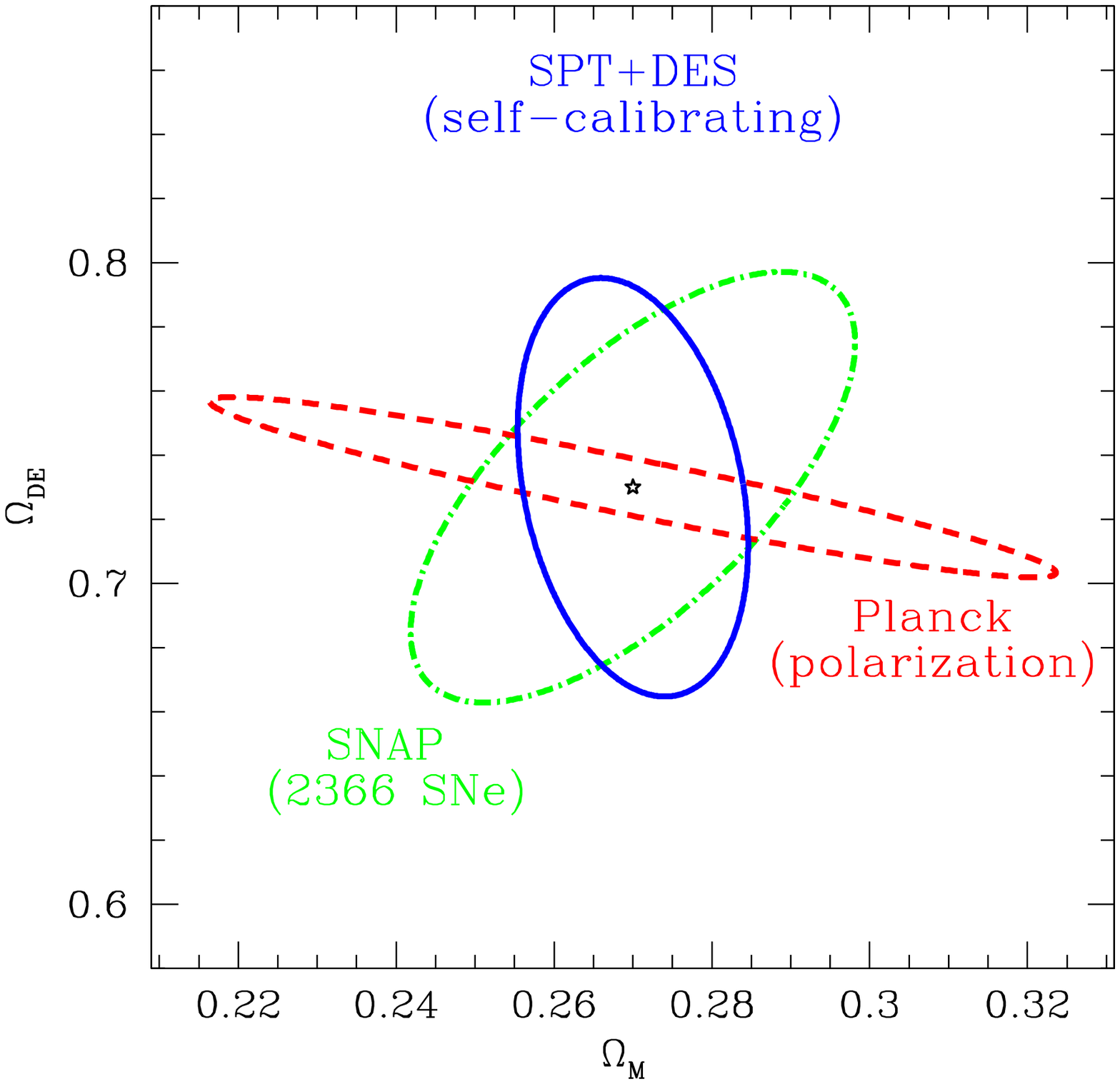}{1.9in}{0}{30}{30}{-190}{-60}
\vskip-2.1in
\plotfiddle{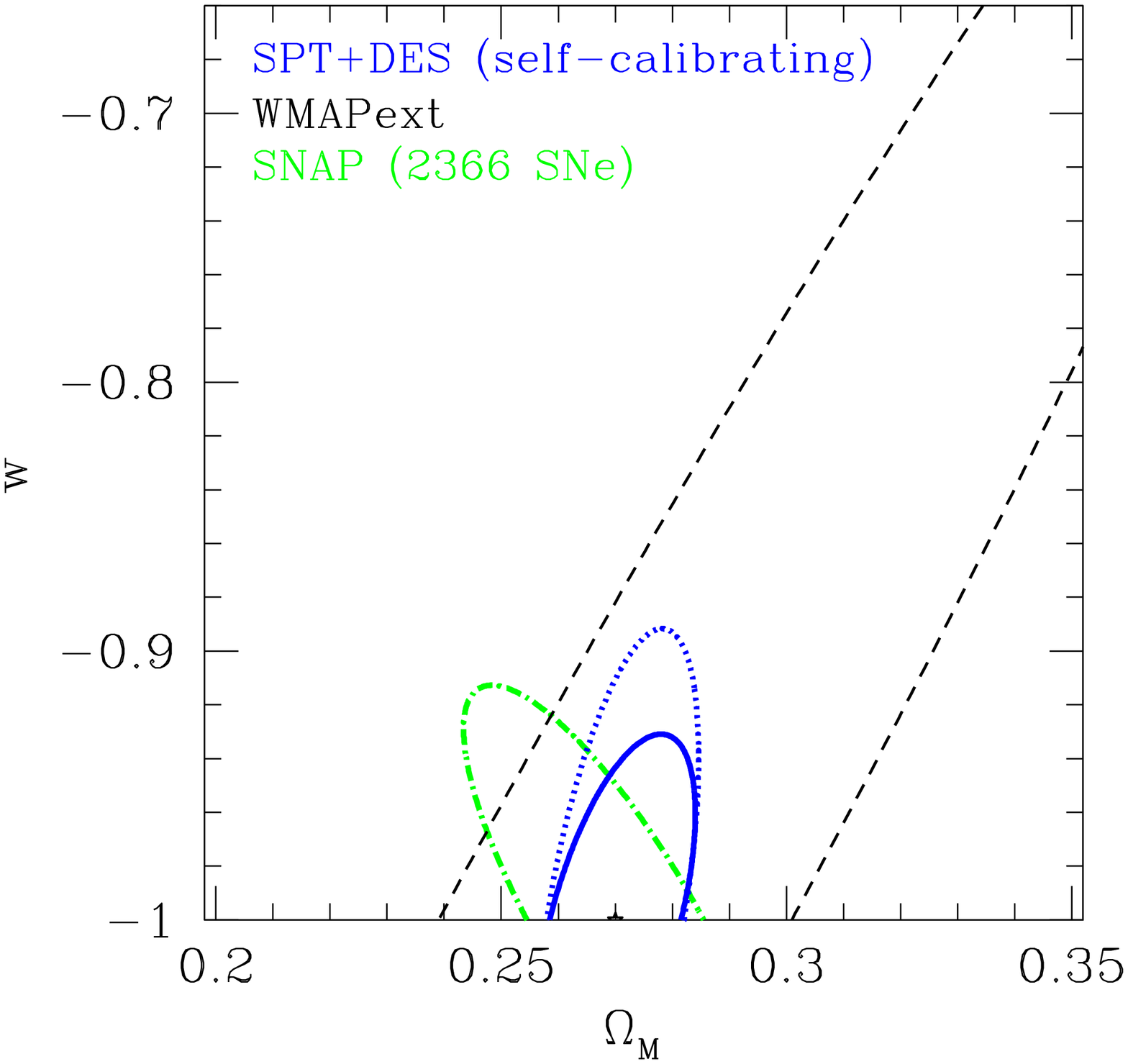}{1.9in}{0}{30}{30}{-10}{-60}
\caption{Forecasts for the geometry constraints (left) from the SPT+DES galaxy cluster survey, the SNAP SNe Ia mission, and the {\it Planck} CMB anisotropy mission together with forecasts for the 
joint $\Omega_M$-\w constraints from the WMAP CMB anisotropy mission, SNAP and SPT+DES.  
Note that the SPT+DES galaxy cluster constraints are competitive with these other leading cosmology experiments; these experiments are highly complementary because each experiment constrains a different combination of cosmological parameters and is subject to different systematics.
\label{fig:spt}}
\end{figure}

The combined SPT+DES cluster survey should provide a strong handle on the nature of the dark energy.  Figure~\ref{fig:spt} contains forecasts for the joint constraints in $w$-$\Omega_M$ space from the cluster redshift distribution, the cluster power spectrum, and 100 mass measurements (each with 30\% 1$\sigma$ accuracy) distributed in mass and extending to $z=1.2$ \citep{majumdar04}.  These forecasts include {\it self--calibration}, which accounts for uncertainties in the mass--observable relation and its evolution.  Cluster finding and masses should arise primarily from the SZE data, and the DES optical data will provide photometric redshifts (with an accuracy of $\delta z\sim0.02$ out to $z\sim1.3$).  The fully marginalized 68\% constraint on constant \w models, is $\delta w=0.05$ (geometry fixed) and $\delta w=0.07$ (geometry freely varying).  In addition, the joint constraints in $\Omega_M-\Omega_E$ space are shown.  For comparison, the constraints expected from SNAP \citep{perlmutter03} and two CMB anisotropy experiments \citep{eisenstein99b,spergel03} are also presented.  It is clear that the self-calibrated galaxy cluster survey constraint is similar in precision to those from other forefront techniques;  in addition, each technique constrains a different combination of cosmological parameters and is subject to different systematic uncertainties, making these experiments highly complementary.

The SPT+DES cluster survey will cover 4000~deg$^2$ of the extragalactic sky south of declination $\delta=-30^\circ$.  Within the concordance cosmological model, it should detect almost 30,000 clusters extending to beyond redshifts $z=1.5$.  The SPT multifrequency SZE survey will be carried out with a new 10~m telescope and 1000--element mm-wave bolometer array \citep{ruhl04}.  The survey is expected to take several south pole observing seasons, and it will commence in February 2007.  The DES multiband photometric survey is currently scheduled to begin surveying in September 2009, and continue for five years.  We expect the first cluster science constraints to appear in late 2010.

An interesting complement to SPT+DES is DUO (PI: Richard Griffiths, CMU), a NASA Small Explorer mission in Phase A study.  If chosen for flight, DUO will execute a two year X--ray survey beginning in 2008 using PN-CCD detectors similar to those operating on XMM--{\it Newton} and mirrors with heritage tracing back to the German Abrixas mission.  The angular resolution in the survey data will be 25\arcsec~FWHM, and this will be quite uniform because of the survey strategy of building up exposure using a large number of different pointings.  This angular resolution is essentially identical to the on--axis resolution of the ROSAT PSPC, which is an earlier detector that has been widely used to find and study clusters out to redshifts beyond $z=1$.  The DUO survey will provide broad band coverage (0.1-10keV) of 6000~deg$^2$ in the north overlapping the Sloan Digital Sky Survey (SDSS), detecting clusters with an X--ray flux at or above $9\times10^{-14}$~erg~s$^{-1}$cm$^{-2}$ [0.5-2keV band]; it will cover 176~deg$^2$ near the south galactic pole to a cluster flux limit of $1.1\times10^{-14}$~erg~s$^{-1}$cm$^{-2}$ in the same band.  The galaxy cluster lgN--lgS is well known to these depths \citep{gioia01}.  Thus, we can say with confidence that the survey will deliver 8000 clusters extending to $z\sim1$ in the north and 1800 clusters extending beyond $z=1$ in the south.  In contrast to the SPT+DES survey, DUO probes a portion of the cluster population that is well understood and employs techniques that have been used for roughly two decades.  Thus, although its cosmological constraints on \w are not quite as impressive as that of SPT+DES (9\% versus 5\% for each experiment on its own), DUO is a far less scientifically risky enterprise.  It offers an ideal dataset for a high yield, large solid angle, self--calibrating galaxy cluster survey.  The availability of X--ray, SZE and optical data over the same survey fields will enable direct cluster distance estimates, which will also be of cosmological value \citep[i.e.][]{reese02}.

\section*{Robust Cosmological Constraints through Self-calibration}

\begin{figure}[!ht] 
\plotfiddle{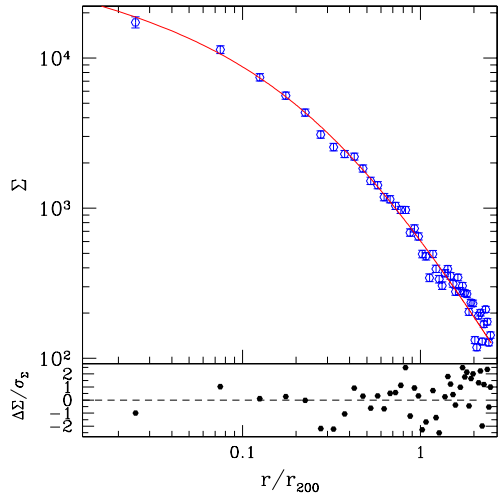}{2.3in}{0}{35}{35}{-110}{-60}
\caption{The galaxy surface density profile for a sample of 93 stacked galaxy clusters studied with 2MASS \citep{lin04a} clearly indicates that the cluster galaxy distribution continues to more than twice the cluster virial radius.  Typically, we adopt a virial radius where the enclosed mean density is a few hundred times higher than the background density.
\label{fig:profile}}
\end{figure}

As noted above, the formation and evolution of dark matter halos is well understood theoretically; it depends on cosmology and the mass of the halos.  Measuring a cluster mass is challenging for many reasons, including the fact that clusters have no outer surface!  Figure~\ref{fig:profile} contains a plot of the projected galaxy distribution \citep{lin04a} of a composite galaxy cluster composed of 93 individual groups and clusters observed with the 2 Micron All Sky Survey (2MASS).  The peak in the X--ray emission defines the cluster center, and the X--ray emission weighted mean temperature is the mass estimator.  The $K$-band galaxy distribution in single clusters is rescaled using an estimate of the virial radius.  It is clear that the galaxy distribution continues with no features well beyond the cluster virial radius $r_{200}$;  the profile is consistent with an NFW model \citep{navarro97} with concentration $c\sim3$.  Structure formation simulations show a similar lack of a virial edge in the dark matter and ICM distributions.  Without a surface to mark the edge of the cluster, one must define a radius for measuring the cluster mass.  Analysis of structure formation simulations (buttressed by theoretical arguments) indicates the formation and evolution of halos can be understood using the simple linear evolution of density perturbations if one adopts a particular cluster mass definition.  The focus is on the so--called virial region, which corresponds to the cluster region where the mean enclosed density is a few hundred times the background or critical densities.  As we demonstrate below, within the framework of self--calibration the exact theoretical definition of the virial region is adopted, and the correlation between that mass and the observable is pulled directly from the survey data.

\begin{figure}[!ht] 
\plotfiddle{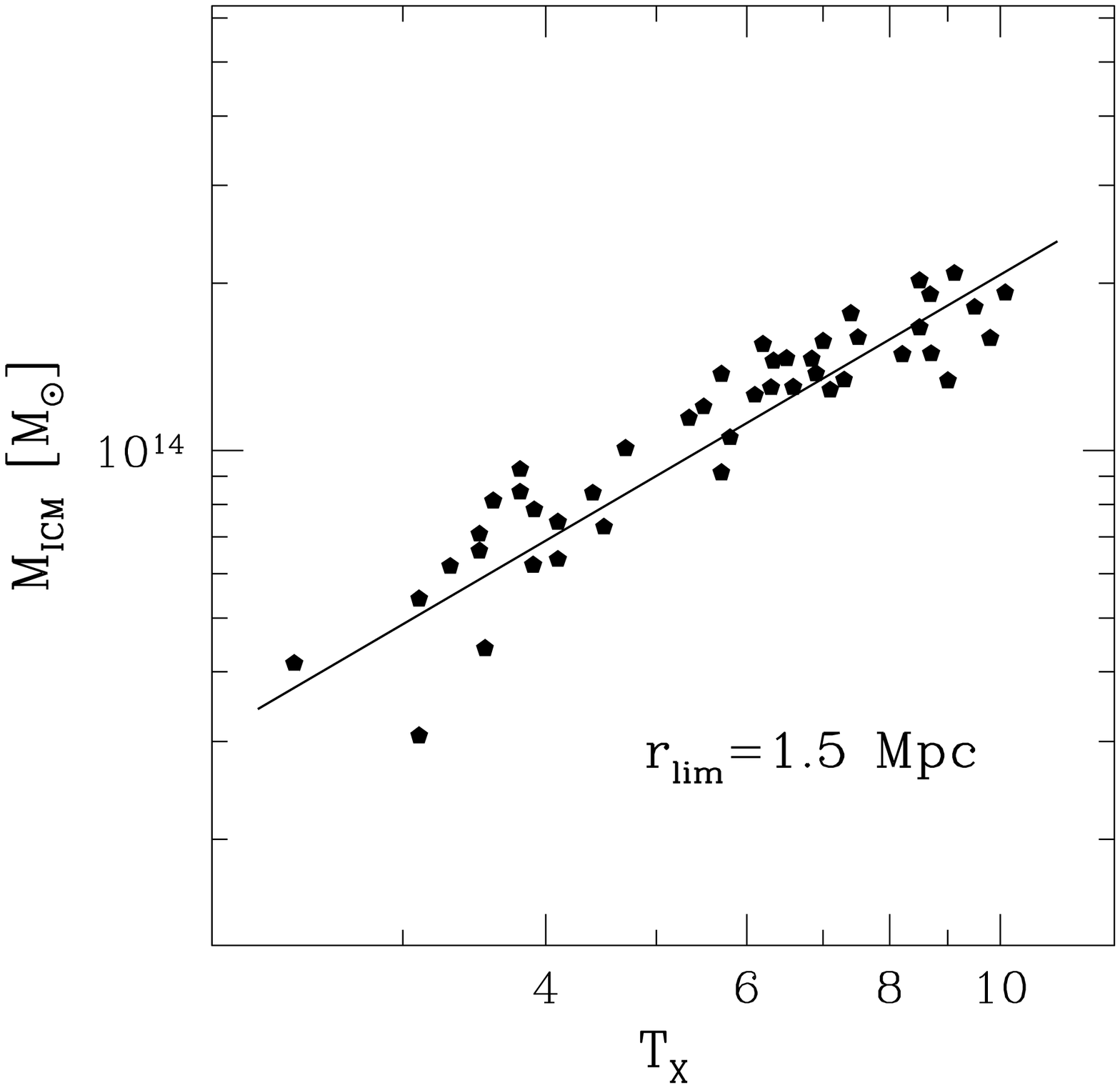}{1.9in}{0}{30}{30}{-180}{-60}
\vskip-2.1in
\plotfiddle{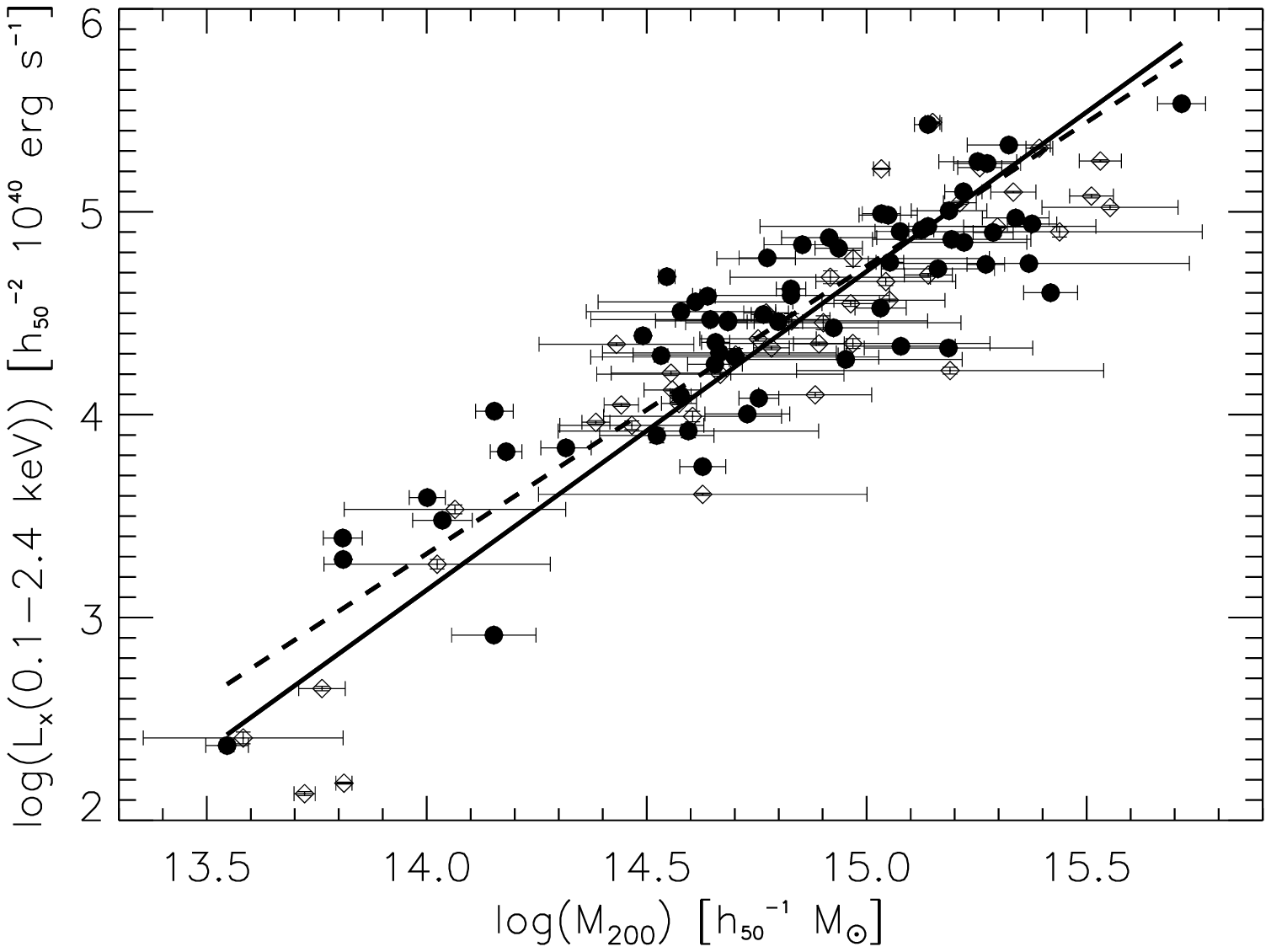}{1.9in}{0}{35}{47}{-30}{-180}
\caption{Observed X-ray scaling relations for clusters indicate that simple observables like the emission weighted mean temperature and X-ray luminosity are good predictors of cluster mass \citep{mohr99,reiprich02}.  This cluster regularity enables the large surveys of $10^4$ clusters required for studies of the dark energy.
\label{fig:scaling}}
\end{figure}

A wide range of observations indicate that clusters exhibit significant regularity through tight scaling relations that exist among independent cluster observables \citep[i.e.][]{mushotzky84,david93}.  The earliest indication of cluster regularity similar in character to the regularity of early type galaxies is in the study of the X--ray isophotal size-- X--ray emission weighted mean temperture relation \citep{mohr97a,mohr00a}.   Understanding the high frequency of cluster substructure \citep[i.e.][]{geller82,dressler88,mohr95,buote96} at the present epoch along with the existence of this regularity (now discussed routinely in the literature) is a very interesting topic in itself, but it lies outside the focus of this presentation \citep[but see][]{ohara04}.  In Figure~\ref{fig:scaling} we show the relation between the emission weighted mean cluster temperature and the ICM mass observed within a fixed metric radius of 1.5~Mpc \citep{mohr99}. This cluster population is a flux limited sample of the 45 brightest clusters in the sky (with many of the most famous merger candidates included).  The scatter is about 20\%, indicating that the emission weighted mean temperature is a predictor of ICM mass with this precision (and accuracy given the small normalization uncertainties).  On the right we show the virial mass--X--ray luminosity relation measured by \citet{reiprich02} using ASCA and ROSAT data.  The scatter here is between 25\% and 30\%, but the inaccuracy may be larger;  it's possible (likelihood is another question) that additional pressure support from turbulence, bulk flow and/or magnetic fields leads to systematic underestimates of these masses, which are determined through the assumption of hydrostatic equilibrium.  The excellent agreement between the WMAP universal baryon fraction and the galaxy cluster baryon fraction (stellar plus ICM mass divided by virial mass) makes it unlikely that there are large scale systematics \citep{lin04a}, but variations in turbulence and bulk flow are almost certainly contributing to the observed scatter in the relation.  Similar scaling relations are available in the optical/NIR, but the scatter in mass at a fixed observable (galaxy number or light within the virial region) is a factor of 3 or so higher than 30\% \citep{kochanek03,yee03,lin03b,lin04a}.  Initial work has been done on SZE scaling relations \citep{cooray99,benson04}, and with the next generation of experiments coming on line, we should know much more within a year.

The existence of these scaling relations and their persistence to intermediate redshifts \citep{mushotzky97,mohr00a,vikhlinin02,ettori04} makes it possible to carry out high yield cluster surveys for precise studies of the expansion history and growth of density fluctuations.  It is impractical to get detailed mass estimates for every cluster in the sample, and, besides, they are not required.  The optical data required to estimate cluster redshifts means that X--ray and SZE surveys will have the data to produce multiple, indepedent mass estimators (ICM properties, galaxy properties, and weak lensing shear for some of the clusters).  A framework for including the mass--uncertainties in cosmological constraints and for exploring the optimal manner for combining all the different cluster data is called self--calibration \citep{majumdar03,hu03b,majumdar04,lima04}.  Within this framework one simply postulates that a relationship between the theoretically relevant halo mass and the cluster observable exists and evolves with redshift.  This relationship and its evolution is then parametrized, and the effects of this additional freedom on cosmological constraints are evaluated using a Fisher technique.  An important component of self--calibration is the combination of multiple observables from the survey, which provides a redundancy that allows for cosmological and cluster mass--observable constraints simultaneously.  These observables include the redshift distribution of detected clusters, the flux distribution of clusters in redshift slices, the clustering of the galaxy clusters, and direct, low accuracy mass measurements of a limited number of clusters.  Ignoring the inaccurate individual mass measurements is tantamount to saying we know nothing about cluster masses.  This is far too pessimistic, because cluster baryon fraction constraints are in good agreement with WMAP CMB anisotropy constraints \citep[i.e.][]{lin04a}; here we postulate that it will be possible to measure 100 cluster masses to 30\% accuracy (1$\sigma$ statistical; these are distributed into ten redshift bins, and by implication we are assuming it is possible to control cluster mass systematic errors to better than the 10\% level).

\begin{figure}[!ht] 
\plotfiddle{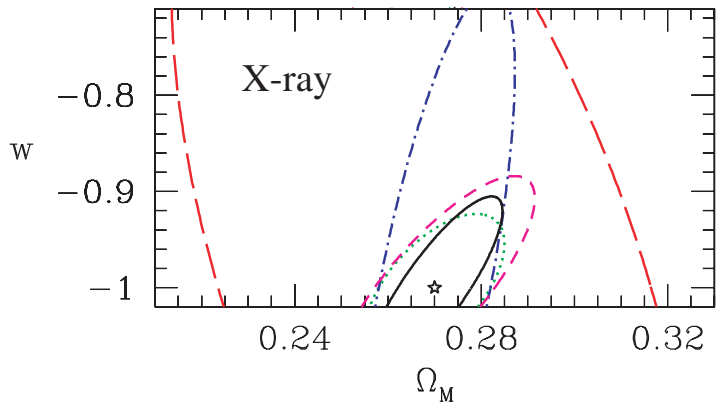}{1.8in}{0}{70}{70}{-160}{-00}
\caption{Forecasts using different combinations of the cluster data from a dedicated X-ray survey mission \citep{majumdar04}:  (a) results from $dN/dz$ that assume perfect knowledge of cluster mass (dotted), (b) self-calibration using $dN/dz$ (long dashed), (c) self-cal using $dN/dz$+$P(k)$ (dash-dot), (d) self-cal using $dN/dz$+mass followup of 100 clusters (dashed), and (e) self-cal using $dN/dz$+$P(k)$+mass followup of 100 clusters (solid).  Precise cosmological constraints can be extracted at the same time that the cluster mass--observable relation and its evolution are determined.
\label{fig:selfcal}}
\end{figure}

Figure~\ref{fig:selfcal} shows the forecasts in $w$--$\Omega_M$ space for a large solid angle X--ray survey, which is analyzed using a self--calibration calculation described in the literature \citep{majumdar04}.  We have modeled cosmology using 7 parameters: the Hubble parameter, the matter density parameter, the baryon density parameter, the dark energy density parameter, the dark energy equation of state parameter \w (assumed to be constant), the initial power spectrum spectral index, and the initial power spectrum normalization.  We adopt the WMAP cosmology with a cosmological constant as our fiducial model.  We model the cluster mass--observable relation and it's evolution as a power law with slope, local normalization and non-standard evolution scaling as $\left(1+z\right)^\gamma$, where $\gamma$ is free to vary.  We then use a Fisher technique together with the redshift distribution $dN/dz$, the clustering of the clusters $P_{cl}(k)$, and 100 mass measurements (ten measurements with 30\% $1\sigma$ accuracy in each of ten bins in the redshift range $0.3<z<1.2$).  We assume crude redshift estimates are available for all clusters and that the Poisson sampling noise in the redshift bins dominates uncertainties in the survey selection function.

The dotted line is the forecast using $dN/dz$ alone but assuming perfect knowledge of cluster masses (i.e. the cluster mass--observable relation and its evolution).  The long dashed line is the self-calibration result using $dN/dz$.  Notice the dramatic weakening in the constraints when the mass--observable relation must be calibrated within the survey.  Adding additional survey information helps enormously:  self--cal $dN/dz$+$P_{cl}(k)$ (dot-dashed), self--cal $dN/dz$+100 inaccurate mass measurements (dashed), and self-cal $dN/dz$+$P_{cl}(k)$+100 inaccurate mass measurements (solid).  The bottom line is that by adding in the cluster power spectrum and 100 inaccurate mass measurements, we can self--calibrate the survey and achieve precision similar to that from the redshift distribution alone when assuming perfect knowledge of the cluster mass--observable relation.  Naturally, the accuracy and robustness are dramatically improved in the self--calibration approach.  Moreover, this approach reserves additional survey information for cross checking the result.  Using the best--fit cosmology and the mass--observable relation and evolution parameters, one can predict the flux distribution of clusters within redshift shells and compare directly to the data (remember that $dN/dz$ is an integral over the flux distribution or mass function and does not use shape information).  If the predicted and observed shapes of the flux distributions (mass functions) are consistent, then this is confirmation that the cosmological constraints are robust.  \citep[For a method that uses the shape of the mass function directly and solves for a separate mass--observable relation within each redshift bin in the survey, please see][]{hu03b}.

Note the importance of the cluster mass measurements even though they are only accurate at the 30\% level (1$\sigma$).  We have adopted 10 redshift bins, and so this mass followup program is statistically equivalent to a single $\sim$10\% measurement (i.e. 30\%/$\sqrt{10}$) in each redshift bin.  Thus, this forecast assumes that systematic uncertainties in cluster masses can be reduced to the 10\% level.  With a combination of weak lensing, deep X--ray, and optical spectroscopic observations, we think this is a reasonable target over the next five years.  Doing even better on the direct mass measurements would allow even stronger dark energy constraints from the survey.

The implications of self--calibration as described above are quite broad.  The following are a few examples.  Uncertainties in the absolute calibration of the observables (perhaps because of fundamental uncertainties in the effective area of the telescope) will just self--calibrate out.  A redshift or mass dependent variation in the AGN contamination of the cluster flux will just self--calibrate out (e.g. a high X--ray contribution to cluster flux at high redshift would be reflected in the non-standard evolution parameter $\gamma$).  Angular filtering to separate cluster SZE flux from the CMB will introduce redshift dependent cluster flux errors, but these would just self--calibrate out.  Systematic photometric redshift uncertainties that vary with redshift would self--calibrate out as non--standard evolution of the mass--observable relation.  The cosmology--dependence of the theoretical halo mass definition is largely irrelevant, because in self--calibration one is solving for the relationship between the theoretical halo mass and the observable statistically.  The only requirement is that the functional form of the non--standard evolution have sufficient freedom.  The comparison of observed and predicted flux functions (mass functions) as a function of redshift will provide a sensitive test of whether the adopted function is adequate.  If there are problems, they will be apparent, and the analysis can be repeated with an adequate non--standard evolution function.

In closing, we note that the forecasts for the SPT+DES surveys shown in Figure~\ref{fig:spt} are calculated in the manner described above; those stronger constraints reflect the higher yield and greater redshift depth of the SPT+DES surveys.  Finally, we emphasize that cluster surveys can be used to constrain far more general models than the constant \w models examined here.  Currently there are two published studies of how cluster surveys can be used to constrain the time variation of the dark energy equation of state parameter \citep{weller02,wang04}.

\acknowledgements{I wish to thank Subha Majumdar and  Zoltan Haiman for many important discussions and fruitful collaborations.  My thanks goes also to the DUO, SPT and DES teams.  This work is supported in part by the NASA grant NAG5-11415 and the NSF award OPP-0130612.}

\bibliographystyle{apj}
\bibliography{cosmology}

\end{document}